\documentclass[twoside,journey]{IEEEtran}

\usepackage{makecell}
\usepackage{hyperref}
\usepackage{array}
\usepackage{caption2}
\usepackage{graphicx,amssymb,amsmath}
\usepackage{multicol}
\usepackage[noadjust]{cite}
\usepackage{setspace}
\usepackage{subfigure}
\usepackage{graphicx}
\usepackage{float}
\usepackage{url}
\usepackage{stfloats}
\usepackage{amsthm,pifont}
\usepackage{flushend}
\usepackage{cases,subeqnarray}
\usepackage{bm,multirow,bigstrut}
\usepackage{amsmath, amsthm, amssymb}
\usepackage{textcomp}
\usepackage{latexsym,bm}
\usepackage{booktabs}
\usepackage{xcolor}
\usepackage{mathtools}
\usepackage{dsfont}
\usepackage{extarrows}
\usepackage{epsfig}
\usepackage{epsfig}
\usepackage{epstopdf}
\usepackage[noend]{algpseudocode}
\usepackage{algorithmicx,algorithm}

\theoremstyle{plain}

\theoremstyle{plain}

\usepackage{amsmath}

\IEEEoverridecommandlockouts
\begin{document}
\title{Adaptive Contextual Caching for Mobile Edge Large Language Model Service}

\author{Guangyuan Liu, Yinqiu Liu, Jiacheng Wang, Hongyang Du$^{*}$, Dusit Niyato,~\IEEEmembership{Fellow,~IEEE}, Jiawen~Kang, and Zehui~Xiong
\thanks{G.~Liu is with the College of Computing and Data Science, the Energy Research Institute @ NTU, Interdisciplinary Graduate Program, Nanyang Technological University, Singapore (e-mail: liug0022@e.ntu.edu.sg).}
\thanks{Y.~Liu, J.~Wang, D. Niyato are with the College of Computing and Data Science, Nanyang Technological University, Singapore (e-mail: yinqiu001@e.ntu.edu.sg, jiacheng.wang@ntu.edu.sg, dniyato@ntu.edu.sg).}
 \thanks{H.~Du is with the Department of Electrical and Electronic Engineering, University of Hong Kong, Hong Kong (e-mail: duhy@eee.hku.hk).}
\thanks{J. Kang is with the School of Automation, Guangdong University of Technology, China. (e-mail: kavinkang@gdut.edu.cn).}
\thanks{Z. Xiong is with the Pillar of Information Systems Technology and Design, Singapore University of Technology and Design, Singapore (e-mail: zehui\_xiong@sutd.edu.sg).}
\thanks{* means the corresponding author}%
}
\maketitle
\vspace{-1cm}
\begin{abstract}
Mobile edge Large Language Model (LLM) deployments face inherent constraints, such as limited computational resources and network bandwidth. Although Retrieval-Augmented Generation (RAG) mitigates some challenges by integrating external knowledge bases, inefficient cache management can still result in high retrieval latency and frequent cache updates. To address these issues, we propose an Adaptive Contextual Caching (ACC) framework that anticipates user needs by proactively caching semantically relevant data for mobile-edge LLMs. ACC utilizes a deep reinforcement learning (DRL) module to refine cache replacement policies, balancing user context, document similarity, and the overhead associated with cache misses. Experimental results demonstrate that ACC increases cache hit rates to over 80\% after only 11 training episodes, outperforming FIFO, LRU, and semantic-only caching while reducing retrieval latency by up to 40\%. In particular, ACC also reduces local caching overhead (i.e., the cost of updating the cache when a miss occurs) by as much as 55\%, enabling scalable, low-latency LLM services in resource-constrained edge environments.
\end{abstract}

\begin{IEEEkeywords}
Retrieval-Augmented Generation (RAG), Caching Strategies, Deep Reinforcement Learning, Proactive Caching
\end{IEEEkeywords}

\section{Introduction}

Large Language Models (LLMs) have significantly advanced natural language processing, offering unprecedented capabilities across a wide range of applications, from customer service chatbots to content generation. These models are highly effective in understanding and generating human-like text, making them invaluable tools in many domains\footnote{https://developer.nvidia.com/blog/tag/large-language-models/}. However, when it comes to deploying LLMs on the mobile edge, they often face performance and latency limitations compared to their cloud-based counterparts. This is due to limited computational resources, limited memory, and bandwidth available on the edge~\cite{qu2024mobile}.

One promising approach to address these challenges in mobile edge LLM deployments is the utilization of knowledge bases~\cite{wu2024stark}. Knowledge bases can take various forms, including knowledge graphs and other types of databases, which provide a structured repository of information that LLMs can leverage to produce more accurate and context-aware responses. By integrating these external data sources, LLMs can mitigate the inherent limitations of model-only knowledge, particularly in dynamic environments where timely information is crucial. Knowledge bases are especially suitable for mobile edge LLM deployments due to their proximity to data sources. This proximity ensures that data remain updated, minimizing the need for data propagation through the Internet, which in turn reduced latency, and the risk of utilizing outdated information. Furthermore, retrieving data locally rather than from distant cloud servers helps preserve privacy by avoiding the transmission of potentially sensitive information over broader networks.

A widely adopted approach to managing and using these knowledge bases in LLMs is Retrieval-Augmented Generation (RAG). RAG is a technique by which an LLM retrieves relevant information from external data sources based on the user prompt and combines it with the original prompt to generate more relevant and informative responses~\cite{jin2024ragcache}. By integrating retrieval mechanisms with generation, RAG can leverage context-specific data effectively, making it particularly well suited for mobile edge scenarios. Mobile edge deployments are typically closer to information sources, such as sensor data, user-generated content, and other localized datasets, allowing for more timely and context-aware responses.

However, employing RAG on the mobile edge introduces challenges related to retrieval efficiency and accuracy. For example, consider a mobile edge LLM used for autonomous vehicle route planning. Using RAG, the model can retrieve the knowledge about the latest traffic data from nearby sources to generate precise, up-to-date routes for specific locations. This localized retrieval capability makes RAG particularly effective at the edge, where timely access to information is crucial. On the other hand, RAG can facilitate task-specific retrieval for legal applications in autonomous driving, such as retrieving regulations governing autonomous vehicle behavior. Compared to dynamic traffic readings, legal regulations are relatively static, meaning that the cache for retrieved content does not need to be updated as frequently as data such as traffic conditions, which change rapidly. Thus, caching becomes a critical strategy for effectively managing data with varying update frequencies. Caching frequently accessed or computationally expensive data closer to the point of use reduces retrieval latency, conserves bandwidth, and optimizes computational resources. For example, caching accessed traffic data or legal information at different frequencies can significantly reduce redundant retrieval operations, thereby improving the overall efficiency of the RAG system.

An effective caching mechanism for mobile edge LLMs involves addressing three crucial aspects: determining what to cache, deciding when to replace cached content, and ensuring that cached information remains relevant and up-to-date. Traditional caching mechanisms, such as Least Recently Used (LRU) and First In, First Out (FIFO), often fail to provide relevant and contextually relevant information in dynamic environments~\cite{singh2024dcache}.

In this paper, we propose an adaptive contextual caching (ACC) framework for RAG in mobile edge LLM services, focusing on optimizing caching mechanisms to enhance the efficiency and responsiveness of the RAG process. The proposed framework is motivated by the need for resource efficiency and adaptability in diverse application domains, ensuring that mobile edge LLMs perform optimally in various use cases. The contributions of this paper are summarized as follows:

\begin{enumerate}
\item We provide a comprehensive overview and comparison of the caching mechanisms used in RAG, discussing their strengths and weaknesses in different scenarios.
    
\item We propose ACC framework that improves the relevance and accuracy of cached information, reducing retrieval latency, and improving overall system performance. Furthermore, we introduce a flexible cache replacement policy that dynamically adjusts to different knowledge domains, optimizing cache management based on the specific characteristics and update frequencies of various LLM applications.
    
\item We develop a deep reinforcement learning (DRL) enabled method to achieve dynamic selection of cache replacement policies, allowing the system to autonomously adapt to changing application requirements and usage patterns.
\end{enumerate}
\section{Contextual Retrieval-Augmented Generation for Mobile-edge LLMs}
\subsection{Challenges of Mobile-edge LLMs}
Traditionally, LLMs such as ChatGPT are hosted on centralized cloud datacenters, enabling them to serve a large number of users simultaneously \cite{bang2023gptcache}. 
However, this centralized approach encounters several critical issues, particularly during peak usage \cite{bang2023gptcache}. 
First, there is a significant increase in latency as data travels from the user's device to distant servers and back. 
Additionally, network congestion can exacerbate these delays, degrading the user experience. 
Furthermore, hosting LLMs on cloud servers raises privacy concerns, as sensitive personal information is transmitted over the Internet, increasing exposure to potential breaches.

Fortunately, mobile-edge LLMs introduce a new paradigm that addresses these issues \cite{jin2024ragcache}. 
By provisioning LLM services closer to the user on the network edge, it is possible to provide customized, low-latency services tailored to individual requirements. 
Despite these advantages, mobile-edge LLMs face significant challenges due to the insufficient hardware capabilities of edge devices, which are less powerful compared to the expansive infrastructure of cloud datacenters. 
For instance, edge servers typically support LLMs with up to a few billion parameters, whereas advanced LLMs like GPT-3.5 boast 175 billion parameters \footnote{https://openai.com/index/gpt-4-research/}. 
This limitation in model size inherently reduces the depth of knowledge that can be learned during the pretraining stage, consequently affecting the LLMs' understanding and generation capabilities\footnote{The generation capability of an LLM refers to its ability to produce coherent, contextually appropriate, and logically consistent content across various modalities, such as text or images, based on its learned knowledge and input prompts.}.
Moreover, data insufficiency presents another significant hurdle. 
Mobile-edge users can hardly collect sufficient and high-quality data necessary for effective LLM training. 
Biased or skewed training data may cause the LLM to generate hallucinations, i.e., generating fact-conflicting answers.
\subsection{Retrieval-Augmented Generation for Mobile-edge LLMs}
\begin{figure*}[t]
\centerline{\includegraphics[width= 1\textwidth]{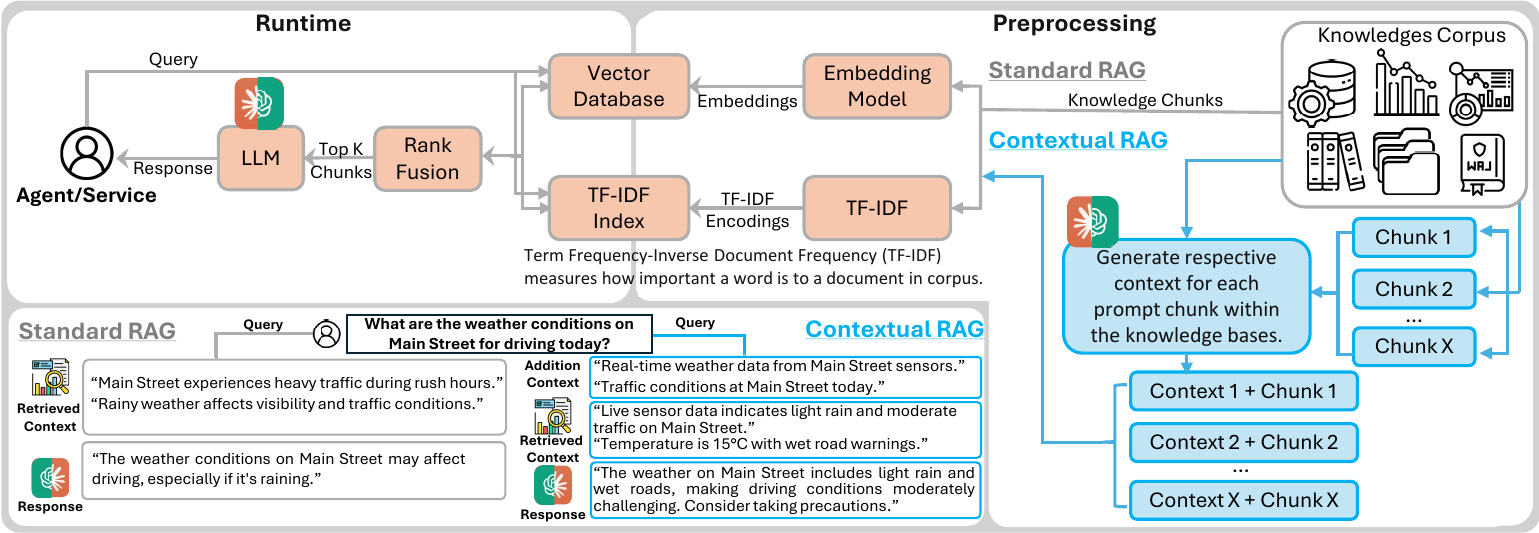}}
\caption{Comparison of Standard RAG and Contextual RAG workflows. Standard RAG retrieves general-purpose knowledge from the knowledge bases to respond to queries, while Contextual RAG augments retrieval with real-time and localized data for enriched responses. The example demonstrates the added value of Contextual RAG in providing actionable insights for a weather-related driving query.}
\label{RAGVSCRAG}
\end{figure*}
RAG represents a transformative approach for enhancing the functionality of mobile-edge LLMs, addressing inherent limitations in computational capacity and resource availability \cite{qu2024mobile}. By seamlessly integrating generative capabilities with retrieval-based precision, RAG dynamically incorporates external, contextually relevant knowledge to optimize performance. This hybrid methodology alleviates the computational burden on LLMs by enabling the retrieval of context-specific information, facilitating a more efficient use of the constrained resources typical of edge environments \cite{jin2024ragcache}.

Unlike in centralized, cloud-hosted LLM systems, RAG-enabled deployments at the mobile edge leverage proximity to localized data sources. This architectural advantage not only reduces latency but also enhances privacy preservation and context-awareness \cite{jin2024ragcache}. These attributes are particularly advantageous in domains such as autonomous driving and healthcare, where the immediacy and relevance of data are critical to operational efficacy. The standard RAG framework, as illustrated in Fig.~\ref{RAGVSCRAG}, includes the following stages~\cite{ganesh2024context}:

\begin{itemize}
\item \textbf{Knowledge Bases Construction}: This foundational step involves the systematic organization and preprocessing of domain-specific or localized data, ensuring that the knowledge base is well-structured and optimized for subsequent embedding and retrieval operations. For mobile-edge LLMs, this often necessitates database tailored to specific edge contexts, such as user-localized data or application-specific knowledge.
\item \textbf{Knowledge Embedding and Chunking}: The prepared bases undergoes transformation into high-dimensional vector representations using embedding models. These embeddings are subsequently partitioned into manageable blocks, facilitating efficient retrieval and processing.
\item \textbf{Knowledge Retrieval}: This stage involves querying the embedded knowledge base to extract the most semantically relevant chunks aligned with the input prompt. Typically, the top-$k$ chunks, determined by a semantic similarity measure, are retrieved. The parameter $k$ can be adjusted to balance retrieval precision and computational overhead.
\item \textbf{Prompt Enrichment}: The retrieved knowledge is integrated into the original user prompt, thereby improving its informational context. This enriched prompt enables the LLM to generate responses that are more nuanced, task-relevant, and accurate.
\item \textbf{Generative Inference}: The enriched prompt is then processed by the LLM. This stage synergistically leverages the model’s intrinsic generative capabilities and the retrieved external knowledge to deliver a comprehensive response.
\end{itemize}

From a system design perspective, the integration of RAG into LLM architectures of mobile edge requires the incorporation of two principal modules. The first is the knowledge bases, which are typically stored in a vector database. Depending on application requirements, this base can be flexibly distributed, with cloud-based repositories maintaining broad, general-purpose knowledge and edge-localized databases housing personalized or domain-specific information. The embedding models are often co-located with the knowledge bases to optimize processing efficiency. 

RAG addresses several challenges intrinsic to mobile-edge LLM deployments:

\begin{itemize}
\item \textbf{Enhance Generation Quality}: By augmenting the limited knowledge scope of edge-deployed LLMs with contextually enriched prompts and external data, RAG markedly improves the relevance and depth of generated content.
\item \textbf{Accomplish Specialized Tasks}: RAG enables LLMs to perform domain-specific or highly specialized tasks by providing access to tailored, high-quality datasets that might otherwise remain inaccessible during the pretraining phase.
\item \textbf{Improve Output Accuracy and Mitigate Hallucination}: By grounding outputs in retrieved, verified knowledge, RAG significantly reduces the likelihood of generating hallucinated or factually erroneous responses, thereby enhancing reliability.
\end{itemize}

\subsection{Enhancing Retrieval-Augmented Generation with Context}

Building upon the foundational RAG framework, Contextual Retrieval-Augmented Generation (CRAG) introduces an adaptive dimension to the knowledge retrieval process \cite{ganesh2024context}. By dynamically tailoring retrieval parameters to the evolving context of user queries, CRAG enhances the specificity and relevance of retrieved information. This approach leverages real-time analysis of prior user interactions and contextual metadata to customize the retrieval process, ensuring alignment with the user’s objectives \cite{ganesh2024context}.

Consider the practical example illustrated in Fig.~\ref{RAGVSCRAG}. Suppose a user in a vehicle queries an LLM about weather conditions for driving on Main Street. A standard RAG system may retrieve broadly relevant information, such as traffic patterns or general weather trends. While accurate, this data lacks the actionable specificity needed for immediate decision-making. In contrast, CRAG refines the retrieval process by incorporating contextual insights—in this case, inferring that the user’s goal is to make informed driving decisions. Consequently, CRAG integrates real-time weather data sourced from local sensors, enabling the LLM to deliver precise and actionable recommendations.

Empirical evaluations of CRAG have demonstrated significant improvements over standard RAG implementations. For instance, CRAG has been shown to reduce information retrieval latency by over 50\% and response generation times by up to 12\%, all while maintaining comparable output quality \cite{ganesh2024context}. Given the centrality of retrieval processes within the RAG paradigm, incorporating advanced caching mechanisms becomes imperative. By caching frequently accessed or computationally expensive data, these systems can further enhance efficiency, reducing latency and ensuring that the most relevant information is readily available. This fusion of contextual retrieval and intelligent caching underpins the scalability and adaptability of mobile-edge LLM solutions.

\section{Caching for Contextual Retrieval-Augmented Generation}

In an RAG system, caching serves as a crucial enhancement mechanism that supports the effective retrieval of information. By storing frequently accessed or computationally expensive data closer to the point of use, caching reduces latency, bandwidth consumption, and computational load, thus facilitating more efficient RAG operations. The following sections review various caching approaches, including traditional methods, semantic and neural techniques, as well as hierarchical and hybrid strategies, providing a comprehensive understanding of how caching enhances RAG performance in mobile edge environments. Using cached context-specific information, such as recent sensor data or domain-specific documents, RAG systems at the edge can ensure faster access to critical knowledge, thus improving system responsiveness and reliability.

\subsection{Caching Systems for Mobile-edge LLMs with RAG}
\begin{figure*}[t]
\centerline{\includegraphics[width= 1\textwidth]{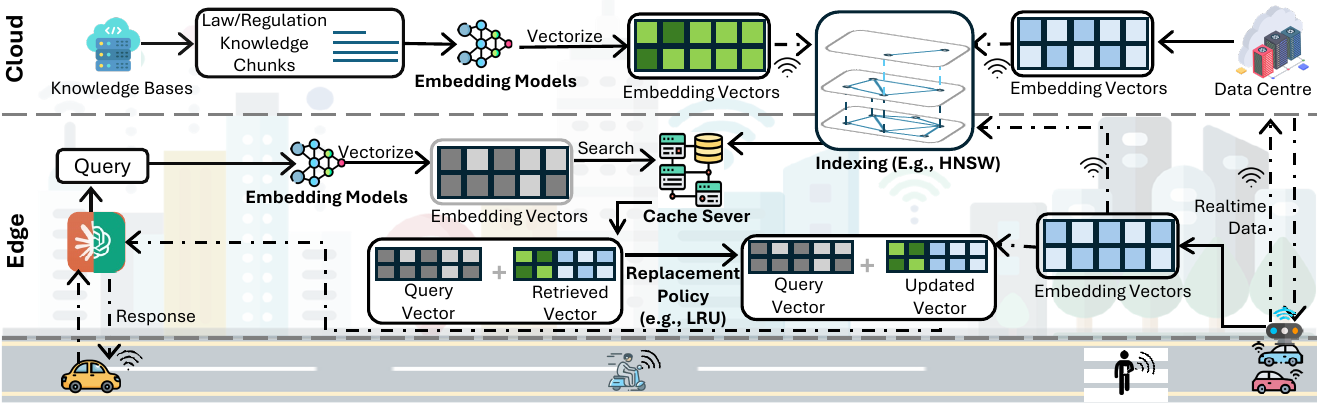}}
\caption{Illustration of key components in a caching system for RAG. The knowledge bases is divided into chunks, vectorized into high-dimensional embeddings via embedding models, and indexed (e.g., using HNSW) for efficient similarity search. Frequently accessed or computationally expensive embeddings are stored in the vector cache, which dynamically replaces expired or non-relevant vectors based on replacement policies. Query embeddings are compared with indexed vectors, and the most relevant contexts are retrieved and combined with the query to form a final prompt for the LLM. }
\label{caching}
\end{figure*}

Effective caching in LLM services relies on several critical components working in concert to efficiently store and retrieve data. These components address key challenges such as determining what data to cache, managing cached data storage, and deciding when to update the cache. As illustrated in Fig.~\ref{caching}, the primary components include embedding models, vector stores with advanced indexing mechanisms, and cache replacement policies.
\subsubsection{\textbf{Embedding Models for Semantic Similarity}}
Embedding models are crucial in determining what data should be cached. They convert information into high-dimensional vector representations that capture semantic relationships, allowing the caching system to store contextually relevant content~\cite{jin2024ragcache}. In LLM services, embeddings of documents, queries, and model outputs enable the system to identify and cache data that is most likely to be reused, thereby reducing inference latency and enhancing computational efficiency~\cite{zhang2019nscaching}. Recent advancements in embedding techniques, such as BERT-based models and Sentence-BERT, have significantly improved the quality of semantic representations, leading to more effective caching decisions \cite{reimers2019sentence}.

\subsubsection{\textbf{Vector Stores and Indexing}}
Vector stores, or vector databases, are essential for managing the high-dimensional embeddings utilized in LLM caching systems. These databases support efficient similarity searches, facilitating rapid retrieval of relevant data~\cite{jin2024ragcache}. Advanced indexing techniques such as Hierarchical Navigable Small World (HNSW) graphs have shown promising results in balancing search speed and accuracy~\cite{malkov2018efficient}. These methods enable LLM caching systems to quickly identify and retrieve the most relevant cached information, even in large-scale deployments. Furthermore, recent work on dynamic indexing strategies allows for real-time updates to the vector store, ensuring that the cached data remains relevant in rapidly changing environments~\cite{guo2020accelerating}.

\subsubsection{\textbf{Cache Replacement Policies}}
Cache replacement policies address the challenge of limited storage by determining when to evict outdated data and replace it with new content. While traditional policies such as LRU and Least Frequently Used (LFU) provide solutions for straightforward scenarios, they often fall short in the complex, context-dependent environment of LLM services~\cite{zhang2019nscaching}.

Advanced policies, such as the Prefix-aware Greedy-Dual-Size-Frequency (PGDSF), have been developed specifically for LLM caching to address these complex requirements~\cite{jin2024ragcache}. PGDSF considers multiple factors, including document order, size, frequency, and recency, ensuring that the most relevant data is retained, particularly in dynamic environments where the relevance of data is continuously evolving.

Recent research has also explored adaptive cache replacement policies that leverage machine learning techniques to dynamically adjust caching decisions based on observed patterns and performance metrics~\cite{narayanan2018deepcache}. These approaches show promise in optimizing cache performance across diverse workloads and usage patterns, which is particularly valuable in the context of mobile edge LLM services where user behavior and data relevance can vary significantly. The effectiveness of LLM caching systems is based on the synergy between embedding models, vector stores with advanced indexing, and sophisticated cache replacement policies~\cite{qu2024mobile}.

In summary, the effectiveness of caching systems for mobile edge LLM RAG relies on the synergy between the above-mentioned components to adapt to the unique challenges of edge computing. Embedding models facilitate the efficient caching of contextually relevant content, enabling edge nodes to provide timely responses without relying on remote servers. Advanced indexing within vector stores ensures rapid retrieval of high-dimensional embeddings, which is crucial for latency-sensitive edge applications. To further address the dynamic nature of mobile edge environments, adaptive caching strategies can respond effectively to conditions such as user mobility and real-time request patterns~\cite{niknia2023edge}. By incorporating such factors into cache replacement decisions, the system can prioritize data that is relevant to the current generation task, thereby improving responsiveness and reducing the need for retrieval from higher-level caches. By effectively leveraging these components, LLM caching systems at the mobile edge can significantly reduce latency, enhance computational efficiency, and provide timely access to relevant information, ultimately improving the scalability and responsiveness of LLM services, particularly in resource-constrained edge environments.

\subsection{Caching Strategies}
\subsubsection{Traditional Caching}

Traditional caching techniques, such as LRU and FIFO have been foundational in various computing contexts including LLM services~\cite{zhang2019nscaching}. These methods are typically used to store data that are accessed frequently, reducing computational burden and improving response time. However, these traditional approaches may fall short in LLM scenarios due to the dynamic and evolving nature of the data being processed. In LLM services, the relevance of cached content is often context-dependent, making it challenging for simple recency-based or frequency-based caching algorithms to provide optimal performance.

\subsubsection{Semantic and Neural Caching}

To address the limitations of traditional caching methods, advanced approaches such as semantic and neural caching have been developed. Semantic caching involves storing data based on its meaning or relevance rather than simply by access patterns. In the context of LLMs, semantic information, such as embeddings, is used to decide what content should be cached. For example, in the scenario of autonomous vehicle congestion prediction, the caching of the embedded data extracted for fused sensor reading instead of the original data in different formats can significantly reduce inference time and improve the relevance of the generated responses by providing immediate access to updated traffic information~\cite{zhang2019nscaching}.

On the other hand, neural caching leverages machine learning models to predict caching decisions. Reinforcement learning-based approaches, such as Deep Q-Networks (DQN), have been employed to learn caching policies that adapt over time to changes in user behavior and query patterns. In recent work~\cite{bai2022knowledge}, knowledge graphs are combined with DQN to provide an additional layer of semantic understanding that further improves caching efficiency by making informed decisions based on the relationships between content elements. This adaptability is particularly valuable in LLM services, where user interactions can vary significantly. Neural caching can help determine which intermediate results or pre-computed outputs should be retained, optimizing the overall performance of the system.

\subsubsection{Hierarchical Caching}

In mobile edge environments, hierarchical caching has emerged as a key strategy to enhance the performance of LLM services. Hierarchical caching involves multiple levels of cache, such as edge nodes, gateways, and cloud servers, each responsible for storing different types of data based on its relevance and access frequency. This multilevel caching strategy is particularly effective in mobile edge deployments, where computational resources are limited, and minimizing latency is critical~\cite{bai2022knowledge}. By caching data at various levels, hierarchical approaches ensure that the most frequently accessed information is available at the edge, while less frequently accessed data is stored in higher-level caches, thus optimizing resource usage and maintaining scalability. For example, in the context of autonomous driving, knowledge related to long-term traffic predictions or regional congestion trends could be cached at a macro base station that covers a larger area. Meanwhile, specific and real-time knowledge about traffic conditions at intersections or upcoming roadblocks could be cached at micro cells closer to those locations.

\section{Contextual Caching: A New Paradigm in the LLM Efficiency of mobile edge LLMs}
The proposed ACC mechanism aims to address the unique challenges of traditional retrieval in RAG for mobile edge LLMs. Unlike traditional caching mechanisms that rely primarily on static, predefined rules, ACC leverages a proactive, context-aware approach to dynamically optimize knowledge storage and retrieval. By contextual analysis, dynamic retrieval, and reinforcement learning, ACC offers a transformative framework to improve the efficiency and responsiveness of mobile edge LLMs.

\subsection{Principles of Adaptive Contextual Caching Framework}
Inspired by contextual retrieval mechanisms~\cite{ganesh2024context}, the ACC framework introduces an intermediate proactive cache server to bridge the gap between the knowledge base and the mobile edge LLM. This server anticipates user needs and adapts its caching strategy dynamically, minimizing retrieval delays and optimizing resource utilization. As illustrated in Fig.~\ref{framework}, the ACC framework contains the following steps:
\begin{figure*}[t]
\centerline{\includegraphics[width= 1\textwidth]{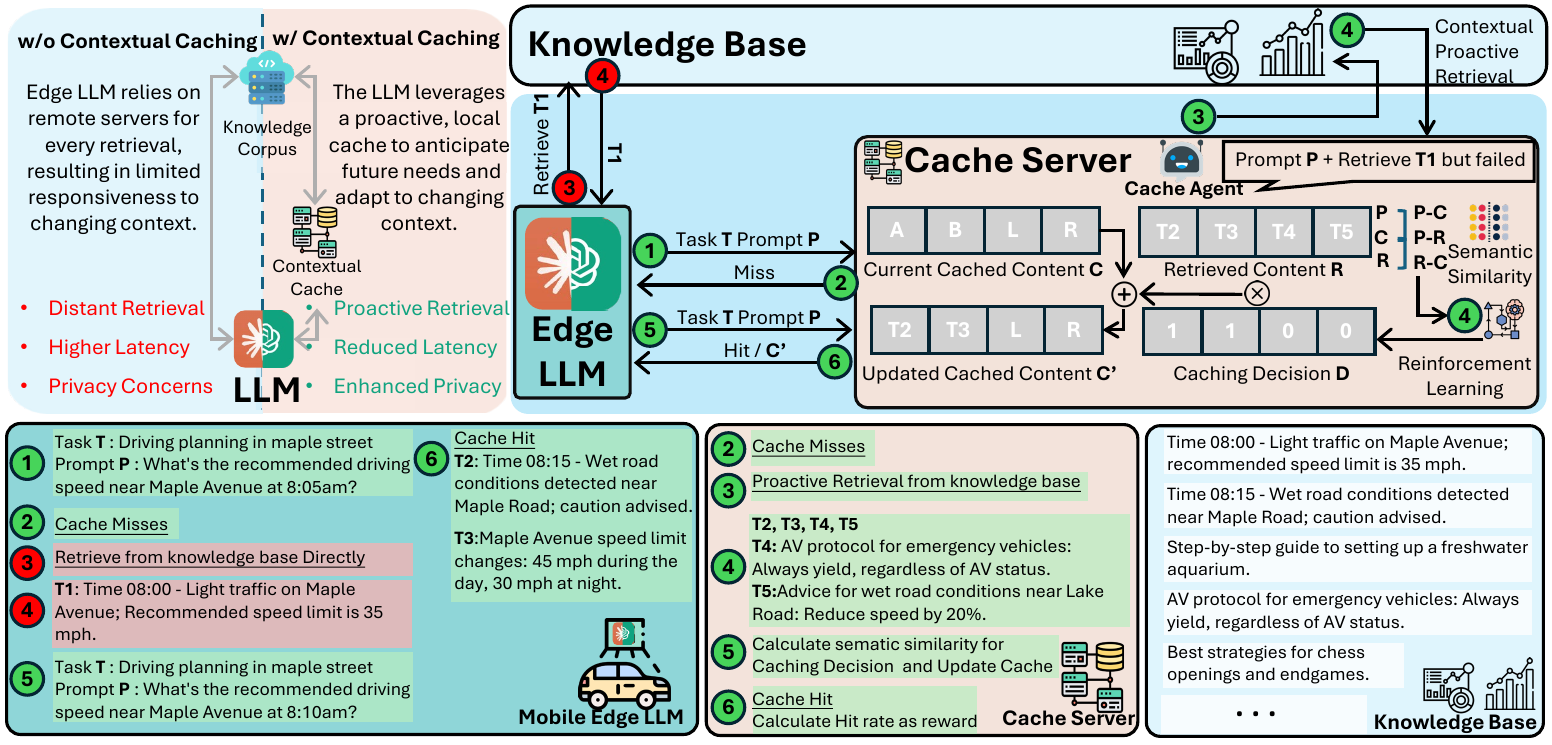}}
\caption{Illustration of the proposed ACC in a mobile-edge LLM scenario, contrasting a conventional retrieval‐only flow with proactive caching approach. Upon receiving a user prompt (\textcircled{1}), the Edge LLM checks the cache server for relevant knowledge. If the cache misses (\textcircled{2}), the system retrieves the needed content (e.g., $T_1$) from the knowledge base (\textcircled{3}). The DRL agent \emph{selectively} updates the cache by deciding whether to store or replace entries (\textcircled{4}). Once updated, subsequent queries can yield a cache hit (\textcircled{5}) without additional retrieval overhead. In the example shown, the mobile unit requests speed guidelines for Maple Avenue and relevant chunks (e.g., $T_2, T_3$) are proactively cached, reducing future latency and overhead.}
\label{framework}
\end{figure*}
\begin{itemize}
\item\textbf{Step 1: Contextual query and initial retrieval.} A mobile edge LLM initiates a retrieval process for task-specific content, i.e., Task $T$, using a prompt $P$. This query triggers a search within the cache to identify relevant knowledge. If the required content is absent, the system records a cache miss.
\item\textbf{Step 2: Cache Miss Handling and Knowledge Base Retrieval.} Upon a cache miss, the mobile edge LLM retrieves the necessary knowledge, denoted as $T_1$, directly from the central knowledge base. Simultaneously, the prompt $P$ and the cache miss information are fed into the proactive caching agent. This agent initiates a contextual analysis to predict additional relevant content, beyond $T_1$, that may be required for future queries.
\item\textbf{Step 3: Contextual analysis and proactive extraction.} The proactive caching agent retrieves a larger set of potentially useful content, denoted as $R$, from the knowledge base. The semantic similarity between the prompt $P$, the current cached content $C$, and the retrieved content $R$ are then calculated as the state information for caching decision. This similarity analysis determines the relevance of $R$ to the ongoing task and informs caching decisions.
\item\textbf{Step 4: Dynamic cache replacement update.} Based on similarity analysis, a DRL model optimizes the cache replacement policy. The DRL model evaluates multiple factors, such as content relevance, update frequency, and storage constraints, to decide whether and how to replace existing cache content. The updated cache, denoted as $C'$, ensures that the most contextually relevant and updated information is readily accessible for subsequent queries.
\item\textbf{Step 5: Reward Feedback for Cache Optimization.} As the edge LLM retrieve the subsequent knowledge from the cache server. The cache hit rate is calculated for consecutive queries related to Task $T$ as a reward function for the DRL model. This feedback loop enables continuous improvement of caching policies, ensuring alignment with evolving user behavior and task requirements.
\item\textbf{Step 6: Iterative Refinement and Retrieval.} For subsequent queries, the mobile edge LLM retrieves task-specific content from the updated cache. If a cache miss occurs, the process repeats from Step 1, enabling iterative refinement of the cache content.
\end{itemize}
\subsection{Implementation in Mobile Edge Environments}
To implement ACC in practical scenarios, the framework employs a modular architecture that integrates seamlessly with existing RAG pipelines. The proactive caching agent acts as an intermediary between the knowledge base and mobile edge LLMs, utilizing vectorized embeddings, advanced indexing, and DRL algorithms to manage cache operations effectively with high scalability and adaptability. The following section explores a case study demonstrating the practical application of ACC in autonomous driving scenarios, highlighting its impact on system performance and user experience.

\begin{figure}[t]
\centerline{\includegraphics[width= 0.48\textwidth]{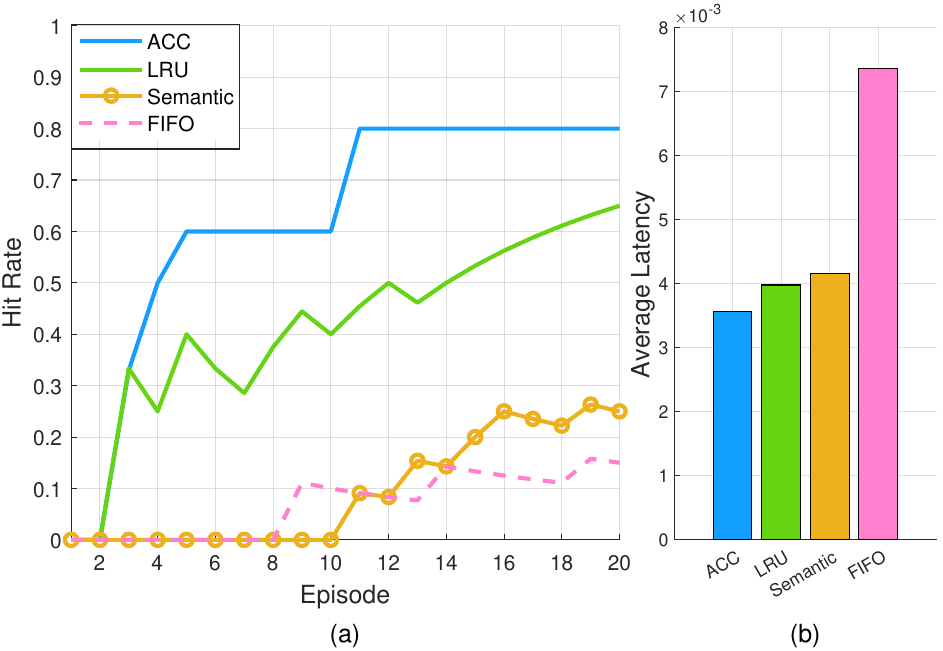}}
\caption{\textbf{(a)} Cache hit rate trends over multiple episodes for different caching strategies. 
    \textbf{(b)} Comparison of the average retrieval latency in seconds. 
    The proposed ACC method achieves higher hit rates and lower latency than baseline approaches.}
\label{fig:hit-rate-latency}
\end{figure}
\begin{figure}[t]
\centerline{\includegraphics[width= 0.48\textwidth]{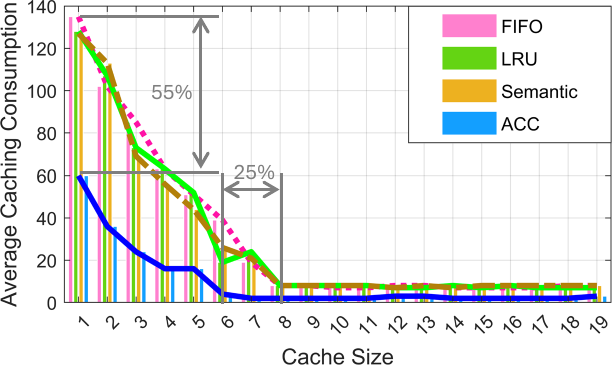}}
\caption{Average caching consumption under varying cache sizes. ACC maintains lower consumption compared to FIFO, LRU, and semantic caching at multiple cache capacities.}
\label{fig:caching-consumption}
\end{figure}

\subsection{Experimental Setup}

We curated a moderate-scale text corpus that intermixes domain-relevant and extraneous content, thereby reflecting realistic scenarios where not all available data directly pertain to the primary application. Each textual segment is first preprocessed (tokenized and normalized) and then partitioned into smaller chunks for fine-grained retrieval. These chunks are converted into compact vector representations using a locally hosted sentence transformer model~\cite{wang2020minilm}, effectively capturing semantic relationships and contextual cues. The resulting embeddings are stored in a vector database to facilitate similarity-based retrieval operations.

Our ACC framework is implemented in a Python environment, where the DRL module is developed using PyTorch. All experiments are conducted on a server equipped with an NVIDIA RTX A5000 GPU (24 GB VRAM). We benchmark the performance of ACC against three baseline caching strategies: 
\begin{itemize} \item \textbf{FIFO}: First-In-First-Out replacement. 
\item \textbf{LRU}: Least Recently Used replacement. 
\item \textbf{Semantic}: Relevance-based replacement~\cite{niknia2023edge}. \end{itemize}

\subsection{Metrics and Results}

\textbf{Hit Rate.} 
Fig.~\ref{fig:hit-rate-latency}(a) presents the changes of cache hit rates across 20 experimental episodes (each episode comprises a fixed number of queries). As the agent interacts with the environment, ACC rapidly converges to a hit rate exceeding 80\%, outperforming all baselines including FIFO, LRU, and purely semantic caching. Clearly, ACC successfully learns an effective replacement strategy through DRL. In comparison, purely semantic caching remains below 30\%. Meanwhile, FIFO struggles to retain relevant data when new chunks arrive, resulting in its lower long-term performance. These results emphasize the importance of dynamic, context-informed decisions in maintaining high cache utilization.

\textbf{Average Latency.}
Fig.~\ref{fig:hit-rate-latency}(b) illustrates the average retrieval latency across the same 20 episodes. Intuitively, the latency of ACC’s mechanisms which include contextual analysis and DRL-based caching policie is higher than baseline methods. However, over multiple episodes, ACC achieves the lowest average latency among the tested methods. This outcome can be attributed to two primary factors. First, ACC’s proactive behavior ensures that relevant knowledge in forthcoming queries is already present in the cache, reducing the frequency of high-latency misses. Second, the cache updates in ACC occur concurrently with knowledge-base retrieval following a miss, which reduces the cost of maintaining the proactive caching strategy. By contrast, LRU and semantic caching display moderate latency and FIFO yield highest average latency due to excessive and frequent unnecessary cache replacements.

\textbf{Local Caching Overhead.}
Fig.~\ref{fig:caching-consumption} illustrates the local caching overhead, measured in terms of the number of data chunks transmitted or updated per miss event. Specifically, each cache miss prompts a retrieval of one or more chunks from the knowledge base, incurring an overhead proportional to the size of the transfer and associated processing. Hence, a higher number of chunks transferred or updated corresponds to a higher caching overhead. In many real-world deployments, excessive overhead in this step can degrade system responsiveness and potentially increase operational costs. As shown in Fig.~\ref{fig:caching-consumption}, ACC reduces overhead by up to 55\% relative to FIFO, LRU, and semantic caching. One explanation lies in ACC’s DRL component, which selectively updates the cache only when necessary.  On the other hand, baseline methods often respond reactively to each miss without a broader strategic view of upcoming queries. Consequently, baseline methods remove cached content prematurely or store knowledge that yields less benefits in subsequent retrievals. From a broader systems perspective, ACC’s lower overhead indicates that the mechanism can scale to larger workloads or more complex edge scenarios without incurring disproportionately high update costs. This advantage becomes increasingly relevant as the size of knowledge bases grows and the frequency of user queries increases.
\section{Future Research Directions}

While the proposed ACC framework demonstrates encouraging performance and robustness in managing contextual caching, there remain several open challenges. These challenges extend beyond the immediate scope of cache replacement policies to encompass a holistic perspective on system architecture, multi-modal data handling, privacy, and resource allocation.

\subsection{Hierarchical Contextual Caching at Scale} The proposed ACC approach has focused on a single-layer caching paradigm. Future work could incorporate hierarchical caching architectures that distribute the caching function across multiple tiers (e.g., user devices, edge servers, and the cloud). Such a multilevel structure would not only facilitate localized retrieval and minimize bandwidth usage, but also enable dynamic load balancing through intelligent offloading of content retrievals between lower-tier and upper-tier caches. Additionally, integrating knowledge of network conditions and user mobility patterns into the caching decisions can further improve retrieval performance in large-scale, geographically distributed networks~\cite{bai2022knowledge}.

\subsection{Cross-Modal Contextual Caching} Contemporary LLMs increasingly operate on multimodal inputs encompassing text, images, audio, and video streams~\cite{li2024generative}. Extending the ACC framework to handle high-dimensional embeddings for these diverse modalities remains an important avenue for future study. By tailoring caching mechanisms to the unique characteristics of each modality, developers can reduce inference delays and ensure timely access to essential context. For instance, in autonomous navigation, integrating high-resolution camera feeds and LiDAR data into the caching pipeline could optimize route planning and environmental perception. Achieving this at scale requires novel indexing structures and fine-tuned replacement policies that preserve essential cross-modal correlations while avoiding cache saturation.

\subsection{Federated and Collaborative Caching} In many real-world mobile edge scenarios, data governance policies and privacy concerns constrain the free exchange of raw data among geographically dispersed edge nodes.  Federated or collaborative caching offers a potential solution by enabling edge nodes to share only learned representations (e.g., embedding vectors or model parameters) instead of raw user data. However, effectively orchestrating these distributed caches—and synchronizing updates in the presence of diverse application demands—poses substantial research challenges. Future investigations may focus on resource allocation strategies and privacy-preserving protocols that allow caches in different network domains to share knowledge securely and adaptively, all while ensuring minimal overhead and maximal quality of service.

\section{Conclusion}
This paper has introduced an ACC framework for mobile-edge LLM services, enhancing RAG through proactive cache management. By leveraging a DRL module, ACC adapts cache replacement policies based on semantic similarity, update frequency, and user behavior. Experimental evaluations demonstrated significantly higher cache hit rates and reduced retrieval latency compared to baseline approaches including FIFO, LRU, and purely semantic caching. The designed mechanism anticipates future data needs in dynamic and resource-constrained edge environments, promoting low-latency responses for domain-specific tasks such as autonomous driving and personalized services. Although our results confirm ACC’s effectiveness, ongoing research will explore hierarchical caching architectures, cross-modal data embedding, and federated learning integration, further expanding the potential of ACC for scalable, efficient mobile-edge LLM deployments.
\bibliographystyle{IEEEtran}
\bibliography{main}

\end{document}